\documentclass[conference]{IEEEtran}
\IEEEoverridecommandlockouts
\usepackage{cite}
\usepackage{amsmath,amssymb,amsfonts}
\usepackage{algorithmic}
\usepackage{enumitem}
\usepackage{graphicx}
\usepackage{textcomp}
\usepackage{xcolor}
\usepackage{float}
\usepackage{amsthm}
\usepackage{algorithm}
\usepackage{mathtools}
\usepackage{tikz}
\usepackage{booktabs}
\usepackage{multirow}
\usepackage[table]{colortbl}

\definecolor{lightblue}{rgb}{0.85, 0.95, 1}

\usepackage{hyperref}

\usepackage{placeins}
\theoremstyle{remark}

\def\BibTeX{{\rm B\kern-.05em{\sc i\kern-.025em b}\kern-.08em
    T\kern-.1667em\lower.7ex\hbox{E}\kern-.125emX}}

\begin{document}

\title{Importance-aware Resource Allocation for Collaborative Task-Oriented Semantic Communication}

\vspace{-2mm}
\author{
Kaiyi Lei$^{1}$, Yuanzhe Peng$^{1}$, Letian Zhang$^{2}$, and Jie Xu$^{1}$\\[2pt]
$^{1}$University of Florida, USA \quad
$^{2}$Middle Tennessee State University, USA\\
Email: \{kaiyi.lei, pengy1, jie.xu\}@ufl.edu, letian.zhang@mtsu.edu
}

\maketitle

\begin{abstract}
Task-oriented semantic communication must allocate scarce radio resources to semantic features under fast fading wireless conditions and strict end-to-end latency budgets.
Existing solutions are either optimization-heavy, leading to prohibitive computational overhead during online operation, or rely on end-to-end retraining procedures together with slowly varying channel assumptions.
We propose iCoTASC (importance-aware Collaborative Task-Oriented Semantic Communication), a hybrid offline--online framework designed for collaborative multi-device semantic communication systems.
iCoTASC leverages attribution-based importance to guide per-dimension embedding selection as a practical communication control signal, models diminishing semantic returns of quantization through a data-driven utility function, and precomputes per-transmitter utility lookup tables offline, which together enable lightweight online scheduling via table lookup and low-complexity refinement under time-varying channels.
The proposed framework supports real-time, channel-adaptive semantic resource allocation in distributed systems without requiring retraining of the underlying task inference model.
\end{abstract}

\begin{IEEEkeywords}
semantic communication, task-oriented communication, resource allocation, importance attribution
\end{IEEEkeywords}

\section{Introduction}

The rapid proliferation of intelligent devices, such as cameras, wearables, and embedded sensors, has transformed modern computing into a distributed, multi-modal ecosystem.
Instead of relying on a single powerful processor, today's systems increasingly operate as teams of interconnected devices that collectively perceive, interpret, and act on their surrounding environment in a coordinated manner \cite{peng2025simac}.
Conventional wireless communication pipelines, grounded in Shannon's separation principle, are designed to faithfully transmit raw data or accurately reconstructed signals over bandwidth-limited links.
Yet, in many emerging applications, the real objective is no longer to preserve raw bits themselves, but to preserve task-relevant meaning that directly supports downstream inference or decision-making.
This shift motivates the emerging paradigm of task-oriented semantic communication, in which transmitters directly encode information that is most informative for maximizing downstream task performance rather than signal-level fidelity \cite{liu2025lightweight}.
By prioritizing semantic utility instead of bit-level fidelity, semantic communication enables more efficient, adaptive, and intelligent interactions among distributed devices and users, particularly under constrained wireless resources and dynamic operating conditions \cite{xie2021deep, wang2024adaptive}.

Recent years have witnessed growing efforts to develop semantic communication architectures and learning-based transceivers \cite{park2025vision, zhang2025seer}.
However, most existing approaches remain primarily limited to single-transmitter scenarios and often assume relatively stable channel conditions together with fixed or pre-determined resource budgets \cite{zhang2024performance, xia2024wireless}.
Such assumptions become increasingly restrictive in practical multi-device use cases such as AR/VR streaming, collaborative perception for autonomous systems, and multi-sensor health monitoring, where task success critically depends on the coordinated and complementary contributions of multiple devices.
Meanwhile, fluctuating bandwidth, latency, and interference conditions demand agile, real-time adaptation of both transmission precision and resource allocation strategies.
Addressing these intertwined challenges calls for a new paradigm, namely \textbf{collaborative task-oriented semantic communication}, which ensures that distributed devices convey not merely raw data, but task-relevant semantic information that directly supports the user's inference objective or operational goal.

\begin{figure}
\centering
  \includegraphics[width=0.45\textwidth, trim=360 210 295 125,clip]{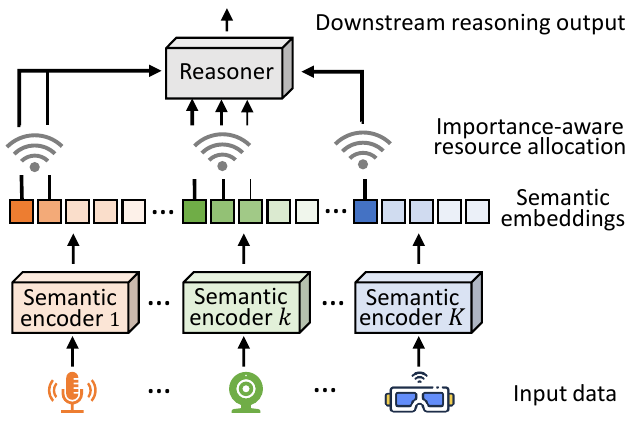}
  \vspace{-3mm}
  \caption{
We propose iCoTASC, an \underline{i}mportance-aware \underline{Co}llaborative \underline{TA}sk-oriented \underline{S}emantic \underline{C}ommunication framework for multi-device systems operating under dynamic wireless conditions.
}
  \label{fig1_intro}
\end{figure}

Designing a collaborative task-oriented semantic communication system introduces several fundamental \textbf{challenges} that extend beyond existing single-transmitter semantic compression.
First, in multi-device settings, the utility of one encoder's representation depends on the embeddings provided by others, because downstream inference relies on the joint, and often complementary, semantic evidence aggregated across devices.
Training each encoder independently therefore risks redundancy, omission of key information, and even implicit overfitting to a single viewpoint or modality, which can reduce robustness under distribution shifts.
Hence, encoders across devices should be jointly trained to produce complementary, synergistic, and non-redundant semantic representations that collectively maximize task performance.
Second, even when semantically rich embeddings are produced, limited bandwidth and time-varying wireless conditions prevent all information from being transmitted, which is particularly problematic for mobile and immersive systems where channel quality can fluctuate rapidly within short time scales.
Therefore, the system must dynamically determine which components of the semantic embeddings to transmit, at what precision, and with what resource allocation, so that it can remain responsive and task-effective under constrained and volatile connectivity.

To address these challenges, we propose iCoTASC, an \underline{i}mportance-aware \underline{Co}llaborative \underline{TA}sk-oriented \underline{S}emantic \underline{C}ommunication framework. 
iCoTASC jointly trains multi-device encoders and extracts dimension-level importance via Integrated Gradients to quantify each embedding component’s contribution to task performance. 
These importance scores serve as criter for runtime embedding selection and quantization. 
We model the impact of quantization using a Weibull-shaped surrogate utility function and formulate a joint optimization problem under per-device bit budgets and a shared RB constraint. 
To enable real-time operation, we decompose the problem into offline per-transmitter utility tabulation and lightweight online greedy scheduling. 
Simulation results show that iCoTASC achieves higher inference accuracy than importance-agnostic baselines under tight bandwidth and fast-varying channels in practical deployment scenarios.

\section{Related Work}
\label{Related Work}

\textbf{Single-User Semantic Communication.}  
Early works integrated deep learning into communication systems to jointly optimize transmitter and receiver models \cite{xie2021deep}. 
Subsequent studies extended this framework to image and language tasks \cite{pokhrel2025harnessing}. 
Unlike these single-user settings, we focus on collaborative multi-device scenarios with shared resource constraints.

\textbf{Multi-User Semantic Communication.}  
Recent studies considered multi-device training and feature fusion \cite{gong2023scalable, peng2026equilibrium, peng2024joint, wei2025task}. 
These works primarily emphasize signal aggregation or joint modeling. 
In contrast, our approach targets importance-aware resource allocation and online scheduling under dynamic channel conditions for real-time distributed inference.

\textbf{Task-Oriented Semantic Communication.}  
Vision-based semantic encoders and large-model approaches have been proposed to maximize inference accuracy \cite{park2025vision, liu2025lightweight}. 
However, resource awareness and fine-grained precision control are often not explicitly incorporated into the communication design.

\textbf{Resource Allocation in Semantic Communication.}  
Adaptive allocation and rate-splitting strategies have been explored to balance semantic fidelity and communication efficiency \cite{qin2024ai}. 
Our work differs by explicitly linking dimension-level semantic importance with offline utility tabulation and lightweight online scheduling.

\section{iCoTASC Framework}
\subsection{System Model}
We consider a system of $K$ transmitters collaboratively delivering task-relevant semantic representations to a shared receiver.
The system operates in discrete time slots.
In each slot, transmitter $k \in [K]$ observes a local input $x_k$ (e.g., a video frame, audio segment, or sensor measurement) and encodes it into a compact latent embedding $h_k = f_k(x_k; \theta_k)$, where $f_k(\cdot; \theta_k)$ denotes the local encoder parameterized by $\theta_k$.
The receiver hosts an inference model $g(\cdot; \phi)$, which fuses the embeddings $(h_1, \ldots, h_K)$ from all transmitters to produce a task-level prediction $\hat{y} = g(h_1, \ldots, h_K; \phi)$.

\textbf{Wireless Transmission Model.}
The underlying wireless system employs orthogonal frequency-division multiple access (OFDMA) with a total of $B$ resource blocks (RBs) available in each time slot.
All RBs share the same time-frequency structure, but their effective payload capacities differ across transmitters due to heterogeneous and time-varying channel conditions.
Specifically, if transmitter $k$ is allocated $b_k$ RBs, its total transmission budget (in bits) equals $b_k c_k$, where $c_k$ denotes the achievable payload per RB determined by the instantaneous channel state.
We assume that all RBs allocated to the same transmitter experience homogeneous channel conditions within a slot, i.e., $c_k$ is constant across those RBs during that slot.
For notational simplicity, we omit the time index, while noting that $c_k$ varies across time and is revealed at the beginning of each slot.

\textbf{Embedding Selection and Quantization.}
Due to limited bandwidth, each transmitter cannot send its full embedding vector in every slot.
Let $H_k = \{1, \ldots, M_k\}$ denote the index set of embedding dimensions for transmitter $k$, where $M_k$ is the embedding size.
In each slot, the system selects a subset of dimensions $S_k \subseteq H_k$ to transmit.
For each selected dimension $m \in S_k$, a quantization level $q_{k,m} \in Q$ is assigned to further compress the data before transmission.
The joint resource allocation, embedding selection, and quantization decisions must satisfy the following bandwidth constraints:
\begin{align}
\scalebox{1}{$
\sum_{m \in S_k} q_{k,m} \leq b_k c_k, \quad \forall k \in [K];
\quad
\sum_{k=1}^{K} b_k \leq B. \label{constraint}
$}
\end{align}
The first constraint enforces that each transmitter's total transmitted bits do not exceed its instantaneous channel budget, while the second ensures that the total number of allocated RBs does not exceed the system-wide RB budget.

\textbf{Impact on Downstream Inference.}
Embedding selection and quantization inevitably lead to perturbed representations $(\hat{h}_1, \ldots, \hat{h}_K)$ received by the server, which can degrade the accuracy of the downstream task $g(\cdot; \phi)$.
Therefore, determining which embedding dimensions to transmit and at what precision is crucial for preserving task performance under dynamic bandwidth constraints.
Our iCoTASC framework addresses this challenge by using importance-aware attribution to guide selection and quantization decisions, and by adapting RB allocation to per-slot channel variations.

\subsection{Importance Attribution via Integrated Gradients}
To enable adaptive and task-effective communication, iCoTASC quantifies the semantic importance of each embedding dimension with respect to the downstream inference task in practice.
The key idea is to measure how much each component of a transmitter's embedding directly influences the receiver's final task output after multi-device fusion.

We adopt Integrated Gradients (IG) \cite{pmlr-v70-sundararajan17a}, a widely used and theoretically grounded attribution method in explainable AI.
IG computes feature importance by integrating the gradient of the model output with respect to its input along a straight-line path from a baseline (uninformative) input to the actual input, thereby providing a principled, sensitivity-based and model-consistent importance score.
For transmitter $k$ with embedding $h_k$, the importance of dimension $m$ is formally defined as
\begin{align}
&\mathrm{IG}_{k,m} =
(h_{k,m} - h_{k,m}')\cdot \nonumber\\
&\int_{0}^{1}
\frac{\partial g(h_1, \ldots, h_k' + \alpha(h_k - h_k'), \ldots, h_K; \phi)}
{\partial h_{k,m}}\, d\alpha,
\end{align}
where $h_k'$ is a baseline embedding (e.g., $h_k' = 0$).

The resulting attribution vector $\mathbf{a}_k = [\mathrm{IG}_{k,1}, \ldots, \mathrm{IG}_{k,M_k}]$ quantifies how much each embedding dimension contributes to the task prediction in the presence of other devices' embeddings.
After normalization, these values form the importance map $\mathbf{w}_k$, which serves as actionable local metadata for runtime transmission control and scheduling decisions.

\subsection{Data-driven Quantization Utility}
To characterize how quantization affects downstream inference performance, we empirically evaluate task accuracy on a validation dataset under varying quantization levels.
The results reveal a clear diminishing-return behavior: as the quantization level steadily increases, the marginal gain in accuracy eventually saturates, indicating that a small number of additional bits on high-importance dimensions can yield disproportionately large gains under tight resource budgets.

\begin{figure}[H]
\centering
  \includegraphics[width=0.43\textwidth, trim=10 0 0 0,clip]{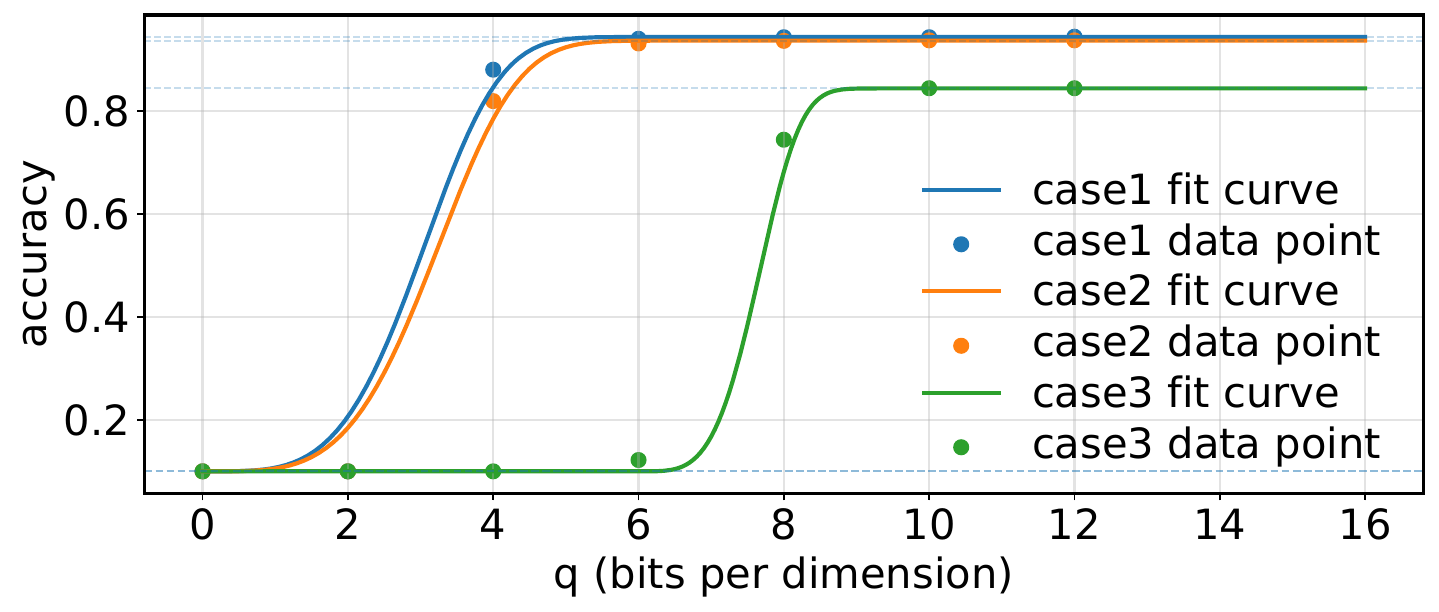}
\caption{Accuracy and fitted utility curves under different quantization levels.
The results for cases~1 and~2 correspond to two and four encoders on the Fashion-MNIST dataset,
while case~3 corresponds to three encoders on the CIFAR-10 dataset.
These results verify that the modified Weibull model closely approximates the quantization--utility relationship.}
  \label{fig:util_curve_fit}
\end{figure}

To model this behavior, we adopt a modified Weibull cumulative distribution function of the following form
\begin{align}
\scalebox{1}{$
u(q) = u_0 + (u_\infty - u_0)\big(1 - e^{-(q/\tau)^4}\big),
$}
\end{align}
where $u_0$ and $u_\infty$ denote the lower and upper bounds of the achievable utility, and $\tau$ controls the overall rate of growth.
These parameters are fitted using validation data offline.
Figure~\ref{fig:util_curve_fit} illustrates the fitted accuracy curves as a function of quantization level across different datasets and numbers of transmitters, demonstrating that the Weibull-shaped surrogate provides a close approximation to the empirically observed quantization--utility relationship.

\vspace{-2mm}

\subsection{Problem Formulation}
Given the importance scores $w_{k,m}$ obtained from Integrated Gradients and the quantization utility function $u(q_{k,m})$, our goal is to maximize the overall semantic utility of the transmitted embeddings under practical bandwidth constraints.
Formally, the joint optimization problem is expressed as
\begin{align}
\max_{b_{1:K},\, S_{1:K},\, q_{1:K}} \quad
& \sum_{k=1}^{K} \sum_{m \in S_k} w_{k,m} u(q_{k,m}) \nonumber\\
\text{s.t.} \quad
& \sum_{m \in S_k} q_{k,m} \le b_k c_k, \quad \forall k \in [K], \nonumber\\
& \sum_{k=1}^{K} b_k \le B, \nonumber\\
& S_k \subseteq H_k,\; q_{k,m} \in Q;\; b_k \in \mathbb{Z}_{\ge 0}. \label{optimization}
\end{align}
This formulation captures the fundamental tradeoff between transmitting semantically important information and adhering to limited per-slot transmission resources under dynamically time-varying channels.

The problem above is combinatorial.
Each transmitter must jointly decide (i) which embedding dimensions to transmit $S_k$, (ii) their quantization levels $q_{k,m}$, and (iii) the number of allocated resource blocks $b_k$.
Because the objective couples discrete selections with a nonlinear utility function and a shared RB constraint across transmitters, the problem is NP-hard.
In particular, even a simplified version with fixed $\{b_k\}$ and linear $u(q)$ reduces to a 0--1 knapsack-type selection problem under a per-transmitter bit budget, which is known to be NP-hard.
Allowing multiple quantization levels and coupling transmitters through the shared RB budget further enlarges the search space, motivating low-complexity algorithms that remain effective under real-time constraints, as developed next.

\section{Hybrid Online-Offline Resource Allocation}
The optimization problem in \eqref{optimization} is NP-hard due to its discrete decision space and coupling between embedding selection, quantization, and bandwidth allocation.
To enable real-time adaptivity, iCoTASC adopts a hybrid offline-online optimization strategy that separates time-invariant and time-varying components.

\subsection{Offline Phase}

The offline phase is executed once per trained model and dataset.
Since the semantic importance weights $w_{k,m}$ and the quantization utility $u(q)$ do not depend on instantaneous channel conditions, we pre-compute quantities that accelerate online decisions.
For transmitter $k$, the semantic utility is:
\begin{align}
U_k(x)
=\max_{S_k,\, q_k}
\{
\sum_{m\in S_k} w_{k,m}u(q_{k,m})
\ \text{s.t.}
\sum_{m\in S_k} q_{k,m}\le x
\},
\end{align}
which is the maximum utility achievable by transmitter $k$ when allocated $x$ bits.
Computing $U_k(x)$ exactly by exhaustive search is prohibitive when the number of embedding dimensions or quantization levels grows.
We therefore adopt an efficient greedy approximation that iteratively assigns one bit at a time to the embedding dimension with the largest current marginal gain.
Let the marginal gain for increasing the quantization of dimension $m$ from $q$ to $q+1$ be
\begin{align}
\Delta_{k,m}(q)=w_{k,m}\big(u(q+1)-u(q)\big).   
\end{align}
At each step, the algorithm selects the dimension with the largest $\Delta_{k,m}(q)$, increments its quantization by one, and repeats until the bit budget $x$ is exhausted.
Pseudocode is given in Algorithm~\ref{alg:greedy}.
When $u(\cdot)$ is nondecreasing and concave, the greedy rule above is optimal for $U_k(x)$.
However, since our $u(\cdot)$ is non-concave, greedy is not guaranteed optimal in general and can be suboptimal.

\begin{algorithm}
\caption{Offline Allocation Algorithm}
\label{alg:greedy}
\begin{algorithmic}[1]
\REQUIRE Importance scores~$w_{k,m}$, quantization set~$\mathcal{Q}$, bit budget~$x$
\ENSURE Offline allocation about ~$\{q_{k,m}\}$ and local utility~$U_k(x)$
\STATE Initialize $q_{k,m}=0$ for all~$m$
\FOR{$total\ bits\ allocated\ t=1$ to $x$}
    \STATE Compute marginal gains when allocating $\Delta_{k,m}(q_{k,m}) =  w_{k,m}[u(q_{k,m}{+}1)-u(q_{k,m})]$
    \STATE Select feature $m^\star = \arg\max\limits_{m}\, \Delta_{k,m}(q_{k,m})$
    \STATE Update $q_{k,m^\star} \leftarrow q_{k,m^\star} {+} 1$
\ENDFOR
\STATE Compute $U_k(x) = \sum_m  w_{k,m}\,u(q_{k,m})$
\end{algorithmic}
\end{algorithm}

\subsection{Online Phase}
During real-time operation, the online phase executes in every time slot to adapt resource allocation and transmission decisions to instantaneous channel conditions.
At the beginning of each slot, each transmitter $k$ observes its current achievable payload $c_k$ (bits per resource block), while the pre-computed offline utility function $U_k(x)$ remains fixed.
The system must then decide how many resource blocks $b_k$ to allocate to each transmitter, subject to the total constraint $\sum_k b_k \le B$.
Pseudocode is given in Algorithm~\ref{alg:online}.

\paragraph{Real-Time Resource Allocation}
Given the offline-computed local utility functions $U_k(x)$ for each transmitter, the corresponding online optimization problem becomes
\begin{align}
\max_{b_1,\ldots,b_K}
\sum_{k=1}^{K} U_k(b_k c_k)
\quad
\text{s.t.} \quad
\sum_{k=1}^{K} b_k \le B, \quad b_k \in \mathbb{Z}_{\ge0}.    
\end{align}
This problem captures how limited resource blocks should be distributed among transmitters to maximize overall semantic utility in practice.
Because $U_k(x)$ is generally non-concave in realistic settings, exact optimization would require enumerating all integer allocations, which is computationally prohibitive for real-time use.
To enable low-latency decision making in practical systems, we employ a greedy marginal allocation strategy: in each iteration, the algorithm computes the marginal utility gain of assigning one additional RB to transmitter $k$,
\begin{align}
\Delta_k(b_k)=U_k((b_k+1)c_k)-U_k(b_k c_k),    
\end{align}
and allocates the next RB to the transmitter with the highest $\Delta_k(b_k)$.
This process repeats until the total budget $B$ is exhausted.
Although not guaranteed to find the global optimum under non-concavity, this greedy allocation yields near-optimal results in practice due to the empirical smoothness and monotonicity of $U_k(x)$.

\begin{algorithm}
\caption{Online RB Scheduling with Precomputed $U_k(\cdot)$}
\label{alg:online}
\begin{algorithmic}[1]
\setlength{\itemsep}{0.3ex}
\REQUIRE Precomputed $U_k(x)$; per-slot payloads $\{c_k\}_{k=1}^K$; total RB budget $B$
\ENSURE RB allocation $\{b_k\}$ and transmitter allocation $\{(S_k^\star, q_k^\star)\}$
\vspace{-2pt}
\STATE Initialize $b_k = 0$ for all $k \in [K]$
\FOR{$total\ RBs\ allocated\ r = 1$ to $B$}
  \STATE $k^\star = \arg\max\limits_{k \in [K]}\ \Big[\, U_k\!\big((b_k{+}1)c_k\big) - U_k\!\big(b_k c_k\big) \,\Big]$
  \STATE \textbf{if} $U_{k^\star}\!\big((b_{k^\star}{+}1)c_{k^\star}\big) - U_{k^\star}\!\big(b_{k^\star} c_{k^\star}\big) \le 0$ \textbf{then break}
  \STATE $b_{k^\star} = b_{k^\star} + 1$
\ENDFOR
\FOR{$k = 1$ to $K$}
  \STATE Retrieve $(S_k^\star, q_k^\star)$ attaining $U_k(b_k c_k)$ from the offline table
\ENDFOR
\STATE \textbf{return} $\{b_k\}$ and $\{(S_k^\star, q_k^\star)\}$
\end{algorithmic}
\end{algorithm}

\paragraph{Local Embedding and Quantization Reconstruction}
After RB allocation, each transmitter reconstructs its detailed transmission plan by consulting the offline-computed utility table $U_k(x)$.
Specifically, transmitter $k$ selects the subset of embedding dimensions and quantization levels $(S_k^*, q_k^*)$ that achieve the value $U_k(b_k c_k)$.
This ensures consistency between offline utility estimation and online scheduling without requiring any per-slot optimization at the transmitter.

\paragraph{Complexity and Practicality}
The greedy online algorithm requires at most $B$ iterations per slot, with each iteration updating the marginal gains of $K$ transmitters.
Using a priority queue, this achieves $\mathcal{O}(B \log K)$ complexity per slot, which is well within real-time constraints for typical wireless systems.
Since all expensive computations (importance attribution, quantization fitting, and $U_k(x)$ estimation) are performed offline, the online phase involves only table lookups and incremental updates, allowing iCoTASC to respond to fast channel fluctuations while maintaining high semantic utility.

\section{Simulation}
\label{sec:exp0}

\subsection{Experimental Setup}

\textbf{Datasets.} We use two public datasets.
\textit{(1) CIFAR-10} contains 50{,}000 training images and 10{,}000 testing images in RGB format ($32{\times}32$) across ten classes.
Each sample is processed by $K{=}3$ encoders, and each encoder produces a 256-dimensional latent representation.
\textit{(2) Fashion-MNIST} contains 60{,}000 training images and 10{,}000 testing images in grayscale format ($28{\times}28$) from ten apparel categories.
We consider configurations with $K{=}2$ and $K{=}4$ encoders.
Each encoder outputs a 256-dimensional latent representation in the two-encoder case and a 128-dimensional latent representation in the four-encoder case.
The reasoner concatenates all received embeddings and applies a classifier to generate predictions.

\textbf{Implementation and Metrics.}
To emulate a resource-constrained edge environment, each run was limited to 16 CPU cores and 32\,GB of system memory.
Training was performed offline in a collaborative manner, where all encoders were jointly optimized with the classifier.
During this stage, embedding importance was estimated using IG, an explainable AI technique that assigns an importance score to each embedding dimension.
The quantization level $q$ was set to multiples of 2 to ensure binary representation.
Quantization parameters were then determined based on the embedding distribution in the validation set.
In the online phase, each channel gain was randomly sampled from the range $[0.1, 0.3]$ during transmission.
The resource block (RB) size was both set to 256 bits for Fashion-MNIST and CIFAR-10, ensuring that results are directly comparable under the same budget setting.
Evaluation metrics include accuracy and computing time.
Accuracy is reported as a function of the RB budget, while computing time refers to the online phase only, excluding offline training.

\textbf{Baselines.} We compare iCoTASC with three baselines.
\textit{Baseline~1:}
At each slot, given the instantaneous $\{c_k\}$ and the RB budget $B$, we use a convex optimization toolbox (CVX) to solve \eqref{optimization}.
Then, the discrete solution is obtained according to $q\in\{0,2,4,\dots,Q_{\max}\}$, and RBs are integerized greedily.
Baseline~1 requires no offline allocation and adapts to the current channel but incurs higher computing time and may be suboptimal due to relaxation and rounding.
\textit{Baseline~2:}
At each RB, the encoder with the highest instantaneous channel gain is selected as the winner.
Two-bit chunks are randomly distributed across that encoder’s dimensions within its capacity and $q_{\max}$.
This method is channel-aware but fails to incorporate importance information into resource allocation.
\textit{Baseline~3:}
This baseline adopts a similar winner selection strategy as Baseline~2 but assigns two-bit chunks with a bias toward lower-$q$ dimensions, giving higher priority to coarsely quantized coordinates.
It achieves broader coverage of dimensions but still ignores the learned importance profile.
In other words, Baseline~2 allocates bits uniformly, whereas Baseline~3 favors lower-$q$ coordinates.

\subsection{Results and Analysis}
This section presents the results of iCoTASC, including (1) accuracy under tight bandwidth budgets and (2) latency and scalability performance.
The results demonstrate the advantages of iCoTASC in achieving higher inference accuracy and lower runtime under dynamic wireless conditions.

\vspace{-2mm}
\begin{figure}[H]
\centering
\begin{minipage}[b]{0.49\linewidth}
    \centering
    \includegraphics[width=\textwidth, trim=10 0 0 0, clip]{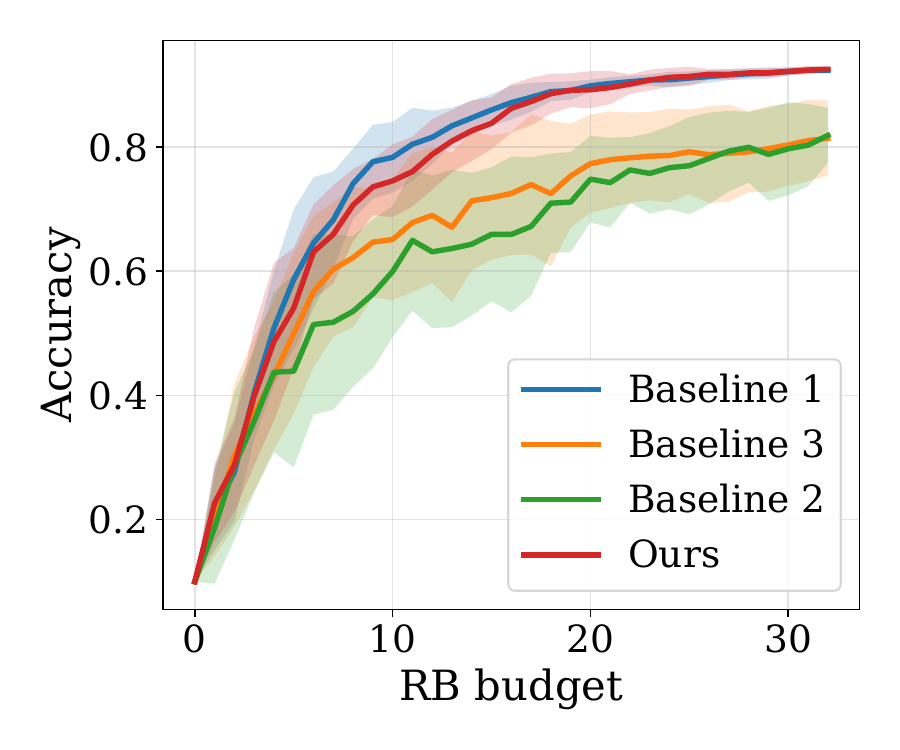}
\vspace{-7mm}
\caption{Accuracy of Fashion-MNIST with two encoders under tight budgets.}
\label{fig:fashion2_acc}
\end{minipage}
\begin{minipage}[b]{0.49\linewidth}
    \centering
    \includegraphics[width=\textwidth, trim=10 0 0 0, clip]{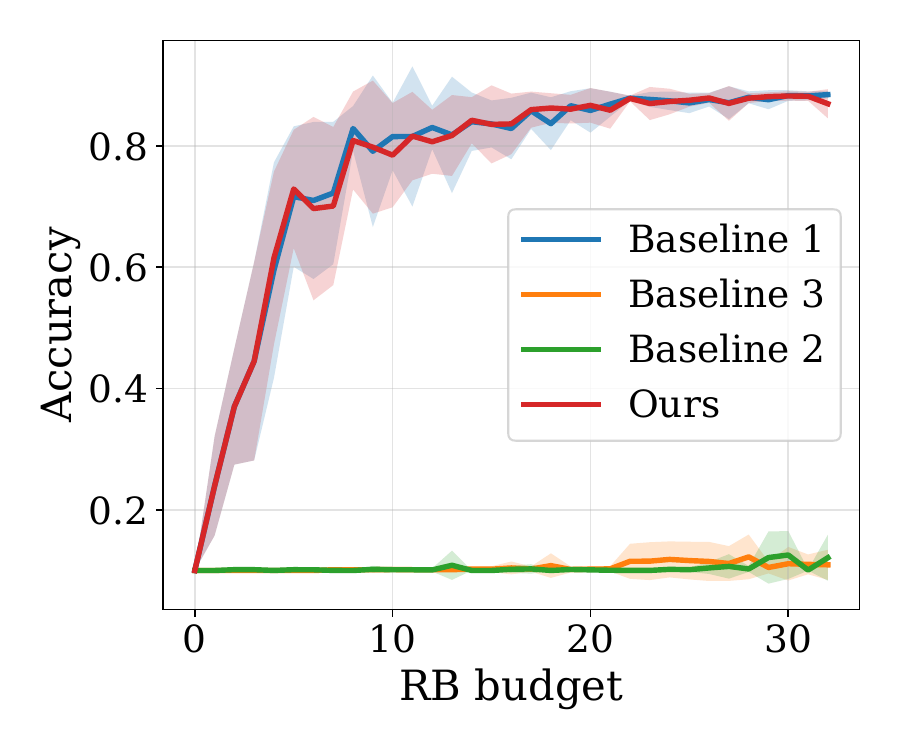}
\vspace{-7mm}
\caption{Accuracy of CIFAR-10 with three encoders under tight budgets.}
\label{fig:cifar_acc}
\end{minipage}
\end{figure}

\textbf{Accuracy under tight budgets.}
Figure~\ref{fig:fashion2_acc} shows the accuracy versus the RB budget on the Fashion-MNIST dataset, while Table~\ref{tab:acc_all} summarizes selected operating points for Fashion-MNIST and CIFAR-10.
Across both datasets, iCoTASC and Baseline~1 consistently outperform the two other baselines in the low-RB region.
On Fashion-MNIST, the accuracy reaches 0.59 with four resource blocks and increases monotonically to 0.85 at thirty two blocks.
On CIFAR-10, as shown in Figure~\ref{fig:cifar_acc}, iCoTASC achieves the highest accuracy at RB~=~32 and maintains that lead through 32 blocks.
The most significant accuracy gains occur within the threshold budget range (e.g., 4 to 32 RBs), where a few additional quantization steps on high-importance dimensions yield substantial improvements in utility.
This observation supports the effectiveness of our quantization design and importance attribution mechanism.
As the available resources increase beyond 8 to 32 RBs, the accuracy improvement gradually slows down while the curves converge, indicating diminishing marginal returns once most dimensions exceed their effective quantization thresholds.

\textbf{Latency and scalability.}
The online scheduling latency ranges from sub-millisecond to a few milliseconds and grows with $B$, as shown in Table~\ref{tab:time_all}.
iCoTASC operates within a similar time-scale envelope as Baselines~2 and~3 and is significantly faster than Baseline~1 (CVX), which must solve a complex allocation problem at each slot followed by rounding.
As the number of resource blocks increases, the runtime grows slightly.
From RB~=~8 to RB~=~16, our method adds 0.266~ms on Fashion-MNIST and 0.318~ms on CIFAR-10 because the latter dataset is more complex.
However, compared with Baseline~1, the increase in computing time remains substantially smaller.
These results align with our motivation of offloading heavy computations (importance extraction, $u(q)$ fitting, and $U_k$ tabulation) to the offline phase while keeping the per-slot process lightweight and limited to priority-queue.

\begin{table}[!htbp]
\centering
\caption{Accuracy across datasets (Fashion-MNIST with 2/4 encoders; CIFAR-10).}
\label{tab:acc_all}

\begingroup
\setlength{\tabcolsep}{3pt}
\renewcommand{\arraystretch}{0.90}
\footnotesize

\textbf{Fashion-MNIST (2 encoders) --- Accuracy}\par\vspace{2pt}
\resizebox{\columnwidth}{!}{%
\begin{tabular}{@{}lccccc@{}}
\toprule
Method & RB=1 & RB=4 & RB=8 & RB=16 & RB=32 \\
\midrule
Baseline1 & 0.237$\pm$0.072 & 0.573$\pm$0.082 & 0.673$\pm$0.083 & 0.762$\pm$0.050 & 0.814$\pm$0.026 \\
Baseline2 & 0.100$\pm$0.000 & 0.181$\pm$0.050 & 0.330$\pm$0.086 & 0.503$\pm$0.126 & 0.656$\pm$0.140 \\
Baseline3 & 0.100$\pm$0.000 & 0.180$\pm$0.051 & 0.314$\pm$0.077 & 0.499$\pm$0.122 & 0.640$\pm$0.137 \\
\rowcolor{lightblue} Ours & 0.216$\pm$0.072 & 0.593$\pm$0.082 & 0.707$\pm$0.070 & 0.796$\pm$0.039 & 0.853$\pm$0.024 \\
\bottomrule
\end{tabular}}

\vspace{6pt}
\textbf{Fashion-MNIST (4 encoders) --- Accuracy}\par\vspace{2pt}
\resizebox{\columnwidth}{!}{%
\begin{tabular}{@{}lccccc@{}}
\toprule
Method & RB=1 & RB=4 & RB=8 & RB=16 & RB=32 \\
\midrule
Baseline1 & 0.189$\pm$0.065 & 0.529$\pm$0.101 & 0.648$\pm$0.082 & 0.753$\pm$0.053 & 0.818$\pm$0.030 \\
Baseline2 & 0.100$\pm$0.000 & 0.165$\pm$0.049 & 0.312$\pm$0.089 & 0.493$\pm$0.122 & 0.644$\pm$0.136 \\
Baseline3 & 0.100$\pm$0.000 & 0.162$\pm$0.050 & 0.294$\pm$0.081 & 0.483$\pm$0.126 & 0.627$\pm$0.137 \\
\rowcolor{lightblue} Ours & 0.194$\pm$0.065 & 0.540$\pm$0.100 & 0.688$\pm$0.070 & 0.789$\pm$0.040 & 0.857$\pm$0.025 \\
\bottomrule
\end{tabular}}

\vspace{6pt}
\textbf{CIFAR-10 --- Accuracy}\par\vspace{2pt}
\resizebox{\columnwidth}{!}{%
\begin{tabular}{@{}lccccc@{}}
\toprule
Method & RB=1 & RB=4 & RB=8 & RB=16 & RB=32 \\
\midrule
Baseline1 & 0.239$\pm$0.082 & 0.595$\pm$0.178 & 0.829$\pm$0.037 & 0.829$\pm$0.051 & 0.885$\pm$0.006 \\
Baseline2 & 0.100$\pm$0.000 & 0.100$\pm$0.000 & 0.100$\pm$0.000 & 0.102$\pm$0.006 & 0.121$\pm$0.038 \\
Baseline3 & 0.100$\pm$0.000 & 0.100$\pm$0.000 & 0.101$\pm$0.004 & 0.104$\pm$0.011 & 0.109$\pm$0.025 \\
\rowcolor{lightblue} Ours & 0.239$\pm$0.082 & 0.615$\pm$0.142 & 0.809$\pm$0.081 & 0.836$\pm$0.050 & 0.870$\pm$0.024 \\
\bottomrule
\end{tabular}}
\endgroup
\end{table}

\begin{table}[!htbp]
\centering
\caption{Computing time (ms) across datasets (Fashion-MNIST with 2/4 encoders; CIFAR-10).}
\label{tab:time_all}

\begingroup
\setlength{\tabcolsep}{3pt}
\renewcommand{\arraystretch}{0.90}
\footnotesize

\textbf{Fashion-MNIST (2 encoders) --- Time (ms)}\par\vspace{2pt}
\resizebox{\columnwidth}{!}{%
\begin{tabular}{@{}lccccc@{}}
\toprule
Method & RB=1 & RB=4 & RB=8 & RB=16 & RB=32 \\
\midrule
Baseline1 & 0.189$\pm$0.018 & 0.300$\pm$0.027 & 0.475$\pm$0.028 & 0.733$\pm$0.058 & 1.382$\pm$0.134 \\
Baseline2 & 0.584$\pm$0.126 & 0.568$\pm$0.106 & 0.556$\pm$0.073 & 0.533$\pm$0.075 & 0.566$\pm$0.087 \\
Baseline3 & 0.581$\pm$0.169 & 0.578$\pm$0.121 & 0.547$\pm$0.075 & 0.529$\pm$0.077 & 0.575$\pm$0.075 \\
\rowcolor{lightblue} Ours & 0.150$\pm$0.023 & 0.263$\pm$0.034 & 0.410$\pm$0.052 & 0.676$\pm$0.046 & 1.219$\pm$0.138 \\
\bottomrule
\end{tabular}}

\vspace{6pt}
\textbf{Fashion-MNIST (4 encoders) --- Time (ms)}\par\vspace{2pt}
\resizebox{\columnwidth}{!}{%
\begin{tabular}{@{}lccccc@{}}
\toprule
Method & RB=1 & RB=4 & RB=8 & RB=16 & RB=32 \\
\midrule
Baseline1 & 0.170$\pm$0.016 & 0.272$\pm$0.018 & 0.449$\pm$0.033 & 0.721$\pm$0.041 & 1.403$\pm$0.093 \\
Baseline2 & 0.571$\pm$0.120 & 0.563$\pm$0.108 & 0.553$\pm$0.074 & 0.546$\pm$0.076 & 0.573$\pm$0.089 \\
Baseline3 & 0.576$\pm$0.140 & 0.568$\pm$0.113 & 0.549$\pm$0.070 & 0.552$\pm$0.070 & 0.586$\pm$0.071 \\
\rowcolor{lightblue} Ours & 0.144$\pm$0.024 & 0.251$\pm$0.021 & 0.394$\pm$0.034 & 0.658$\pm$0.038 & 1.252$\pm$0.093 \\
\bottomrule
\end{tabular}}

\vspace{6pt}
\textbf{CIFAR-10 --- Time (ms)}\par\vspace{2pt}
\resizebox{\columnwidth}{!}{%
\begin{tabular}{@{}lccccc@{}}
\toprule
Method & RB=1 & RB=4 & RB=8 & RB=16 & RB=32 \\
\midrule
Baseline1 & 0.596$\pm$0.014 & 2.171$\pm$0.174 & 3.958$\pm$0.169 & 7.727$\pm$0.092 & 15.900$\pm$0.410 \\
Baseline2 & 0.070$\pm$0.002 & 0.385$\pm$0.107 & 0.799$\pm$0.183 & 1.183$\pm$0.198 & 1.954$\pm$0.091 \\
Baseline3 & 0.073$\pm$0.002 & 0.404$\pm$0.116 & 0.723$\pm$0.097 & 1.162$\pm$0.088 & 1.991$\pm$0.040 \\
\rowcolor{lightblue} Ours & 0.035$\pm$0.001 & 0.124$\pm$0.039 & 0.385$\pm$0.020 & 0.703$\pm$0.098 & 1.554$\pm$0.183 \\
\bottomrule
\end{tabular}}
\endgroup
\end{table}

\textbf{Impact of $K$.}
We further investigate the impact of the number of encoders $K$.
We present results on the Fashion-MNIST dataset and observe similar trends on CIFAR-10.
As shown in Table~\ref{tab:acc_all}, under small RB budgets, the accuracy with the configuration of $K{=}2$ slightly outperforms that with that of $K{=}4$ from four to nine resource blocks.
The maximum accuracy gap is approximately 0.05.
As the RB budget increases, this gap gradually narrows and largely disappears around ten RBs, reflecting the strategy of prioritizing early coverage before refining precision.

\section{Conclusion}
\label{sec:conclusion}

In conclusion, we present iCoTASC, an importance-aware collaborative task-oriented semantic communication framework specifically designed for multi-device systems operating under dynamic wireless conditions. iCoTASC enables multiple transmitters to jointly deliver task-relevant information by integrating explainability-based importance attribution with a data-driven quantization utility model for embedding selection and adaptive precision control. We propose a hybrid offline and online algorithm that precomputes transmitter utilities and performs efficient real-time resource allocation based on instantaneous channel states. The proposed framework is model-agnostic and readily scalable across different tasks. Experimental results demonstrate that iCoTASC consistently achieves higher inference accuracy than the baselines under tight bandwidth and time-varying channel conditions, clearly highlighting the effectiveness of importance-aware resource allocation for scalable distributed intelligence systems.

\section*{Acknowledgment}
The work of Y. Peng, and J. Xu was supported in part by the U.S. National Science Foundation under Awards No. 2515982, 2505381, 2433886. The work of L. Zhang was supported in part by the U.S. National Science Foundation under Award No. 2348279 and the MTSU Stark Land Project.

\bibliographystyle{IEEEtran}
\bibliography{refs}

\end{document}